\documentclass[aps,prb,twocolumn,twoside,floatfix,amsmath,showpacs,superscriptaddress]{revtex4-1}

\usepackage{amssymb}
\usepackage{graphicx}
\usepackage{color}
\usepackage[normalem]{ulem}

\begin{document}


\newcommand{\LaPG}{LaPt$_4$Ge$_{12}$}
\newcommand{\PrPG}{PrPt$_{4}$Ge$_{12}$}

\newcommand{\sovert}[1]{\ensuremath{#1\,\mu\textnormal{V K}^{-2}}}
\newcommand{\resist}[1]{\ensuremath{#1\,\mu\Omega\textnormal{cm}}}
\newcommand{\mJmolK}[1]{\ensuremath{#1\,\textnormal{mJ mol}^{-1} \textnormal{K}^{-2}}}
\newcommand{\lorenzunits}{\ensuremath{\,\textnormal{W \Omega K}^{-2}}}

\newcommand{\ie}{{\em i.e.}}
\newcommand{\eg}{{\em e.g.}}
\newcommand{\cf}{{\em cf.}}
\newcommand{\Eg}{{\em E.g.}}

\newcommand{\kB}{{\ensuremath{k_{\mathrm{B}}}}}
\newcommand{\Tc}{\ensuremath{T_{\mathrm{c}}}}
\newcommand{\Hcc}{\ensuremath{H_{\mathrm{c2}}}}
\newcommand{\kaSC}[2]{\ensuremath{\kappa_\mathrm{#1}^\mathrm{#2}}}
\newcommand{\ka}{\kappa}

\newcommand{\replace}[2]{\sout{#1} \textcolor{red}{#2}}
\newcommand{\tred}[1]{\textcolor{red}{#1}}
\newcommand{\tblue}[1]{\textcolor{blue}{#1}}

\title[]{Superconducting gap structure of the skutterudite \LaPG\ probed by specific heat and thermal transport}

\author{H. Pfau}
\email{Present address: Stanford Institute for Materials and Energy Sciences, SLAC National Accelerator Laboratory and Stanford University, 2575 Sand Hill Road, Menlo Park, California 94025, USA, hpfau@stanford.edu}
\affiliation{Max Planck Institute for Chemical Physics of Solids, 01187 Dresden, Germany}
\author{M. Nicklas}
\email{nicklas@cpfs.mpg.de}
\affiliation{Max Planck Institute for Chemical Physics of Solids, 01187 Dresden, Germany}
\author{U. Stockert}
\affiliation{Max Planck Institute for Chemical Physics of Solids, 01187 Dresden, Germany}
\author{R. Gumeniuk}
\affiliation{Max Planck Institute for Chemical Physics of Solids, 01187 Dresden, Germany}
\affiliation{Institut f{\"u}r Experimentelle Physik, TU Bergakademie Freiberg, 09596 Freiberg, Germany}
\author{W. Schnelle}

\affiliation{Max Planck Institute for Chemical Physics of Solids, 01187 Dresden, Germany}
\author{A. Leithe-Jasper}
\author{Y. Grin}
\author{F. Steglich}
\affiliation{Max Planck Institute for Chemical Physics of Solids, 01187 Dresden, Germany}


\begin{abstract}

We investigated the superconducting order parameter of the filled skutterudite \LaPG, with a transition temperature of $T_c = 8.3$\,K. To this end, we performed temperature and magnetic-field dependent specific-heat and thermal-conductivity measurements. All data are compatible with a single superconducting $s$-wave gap. However, a multiband scenario cannot be ruled out. The results are discussed in the context of previous studies on the substitution series  Pr$_{1-x}$La$_{x}$Pt$_4$Ge$_{12}$. They suggest compatible order parameters for the two end compounds \LaPG\ and \PrPG. This is not consistent with a single $s$-wave gap in \LaPG\ considering previous reports of unconventional and/or multiband superconductivity in \PrPG.

\end{abstract}

\maketitle

\section{Introduction}

The pairing mechanism of a superconductor determines the symmetry of its order parameter, which in turn is connected to the symmetry of the superconducting gap. Superconductors, whose averaged order parameter over the entire Fermi surface yields zero, are called unconventional. They attract much interest \cite{white_2015} in modern condensed matter physics because a description of the underlying physics has to go beyond the standard Bardeen-Cooper-Schrieffer (BCS) theory for an $s$-wave order parameter. Their gap contains nodes, whose existence and position can be detected by a variety of experimental probes. Often, only a combination of results from different probes allows one to draw a conclusive picture. The search and study of unconventional superconductors was triggered by the discovery of heavy-fermion superconductivity \cite{steglich_1979}, the high-$T_c$ cuprates \cite{bednorz_1986}, and organic superconductors \cite{jerome_1980}. By now, a lot more materials are believed to be unconventional superconductors. 

In this context, the filled-skutterudite compounds $MT_4X_{12}$ ($M=$ electropositive metal, $T=$ transition metal, and $X=$ usually a pnictogen) attracted much attention with the discovery of PrOs$_4$Sb$_{12}$, which is the first Pr-based heavy-fermion superconductor and believed to be of unconventional type \cite{bauer_2002,aoki_2005}. It exhibits exotic properties probably connected to the quadrupole degrees of freedom \cite{aoki_2005}. The skutterudite family $M\mathrm{Pt}_4\mathrm{Ge}_{12}$ ($M =$ Sr, Ba, La, Pr, Th) with a Pt-Ge framework are also superconductors  \cite{bauer_2007,gumeniuk_2008,bauer_2008,kaczorowski_2008,jeon_2016}. The two members \PrPG~and \LaPG~ show superconductivity at relatively high transition temperatures compared to other skutterudites, namely $T_c = 7.9\,\mathrm{K}$ and 
$8.3\,\mathrm{K}$ \cite{gumeniuk_2008}. 

\PrPG~seems to be a good candidate for unconventional superconductivity. It is considered to be a moderately strong-coupling superconductor from the large specific-heat jump compared to the BCS value \cite{gumeniuk_2008}. There are indications of point nodes from NMR \cite{kanetake_2010}, specific heat, and penetration depth \cite{zhang_2013}. Furthermore, $\mu$SR measurements detected a time-reversal symmetry breaking below $T_c$ \cite{maisuradze_2010,zhang_2015}. Additionally, a number of investigations including photoemission, magnetization, penetration depth, and specific heat revealed multiband superconductivity in this compound \cite{nakamura_2012,chandra_2012,zhang_2013,huang_2014,singh_2016}.

The continuous evolution of $T_c$ across the doping series (Pr$_{1-x}$La$_x$)Pt$_4$Ge$_{12}$ suggests compatible order parameters of the end members \PrPG~and \LaPG\ \cite{maisuradze_2010}. However, the few existing investigations on \LaPG~point towards a single isotropic gap: the specific-heat jump $\Delta C/\gamma_\mathrm{n}T_\mathrm{c}$ is only slightly above the BCS value suggesting a weaker coupling than in \PrPG\ \cite{gumeniuk_2008}. NMR and photoelectron-spectroscopy results for \LaPG\ could be best explained by a single isotropic gap \cite{toda_2008,nakamura_2012}. Additionally, no time-reversal symmetry breaking is observed in \LaPG\ by $\mu$SR \cite{maisuradze_2010}. Only one recent study using $\mu$SR and tunnel diode spectroscopy for penetration-depth measurements reports indications of multiband superconductivity in \LaPG\ \cite{zhang_2015_2}.

In order to develop a conclusive picture, it is necessary to shed more light on the superconducting properties of \LaPG. In particular, it is important to clarify, if \LaPG\ is a multiband superconductor. To this end, we performed specific-heat measurements at temperatures down to 0.4\,K for different magnetic fields. While the temperature dependence at zero field can be described with a single superconducting $s$-wave gap, the field dependence of the specific-heat $\gamma_0$-coefficient shows sub-linear behavior, which is discussed in terms of Fermi-surface anisotropies. 
Additionally, we use the specific-heat results to analyze the data of our detailed thermal-conductivity study with a focus on the behavior below 1\,K. Our temperature- and field-dependent measurements are compatible with a single-gap $s$-wave state.

\section{Methods}

For the synthesis of single crystals of \LaPG, a bulk sample of mass 2\,g was placed in a glassy-carbon crucible, sealed in a Ta tube and enclosed in a quartz ampoule. The ampoule was heated up to 870\,$^{\circ}$C within 5\,h and kept at this temperature for 10\,h. Then it was cooled down to 845\,$^{\circ}$C and held for 30 days and furnace-cooled. Single crystals of 1-4\,mm size were mechanically extracted from the sample \cite{gumeniuk_2010}.

We investigated three high-quality single crystals of \LaPG\ selected from the same batch with a residual resistivity ratio (RRR) of 17 and a residual resistivity $\rho_0$ of \resist{3.8}.
Crystal \#1 with a mass of $m=14.7$\,mg and crystal \#2 with 67.2\,mg were used for specific-heat measurements by a relaxation method (Heat Capacity Option in a PPMS by Quantum Design). On crystal \#3 we performed transport measurements with a standard two-thermometer-one-heater technique. For this purpose, we cut it first into a rectangular bar along the principal cubic crystal axes with a cross-sectional area of 0.492\,mm$^2$ and a contact distance of 1.08\,mm. This arrangement was used for thermal-conductivity measurements in a PPMS (Thermal Transport Option) above 2\,K and in zero magnetic field. Afterwards, a long plate was cut from crystal \#3 again along the principal cubic crystal axes with the dimensions $(0.04\times0.50\times1.92)\,\mathrm{mm^3}$. This sample \#3a was used to measure thermal conductivity both as a function of temperature and isothermally as function of magnetic field for temperatures below 1\,K and for magnetic fields of $0\leq H \leq 2$\,T. The heat current $\boldsymbol{j}$ was applied along the same direction as for the PPMS measurement. A superconducting split-coil magnet generated a magnetic field $\boldsymbol{H}\perp \boldsymbol{j}$. 


\section{Results}

\subsection{$T$-dependence of the specific heat}

\begin{figure}[tbh]
  \begin{center}
    \includegraphics[width=0.8\linewidth]{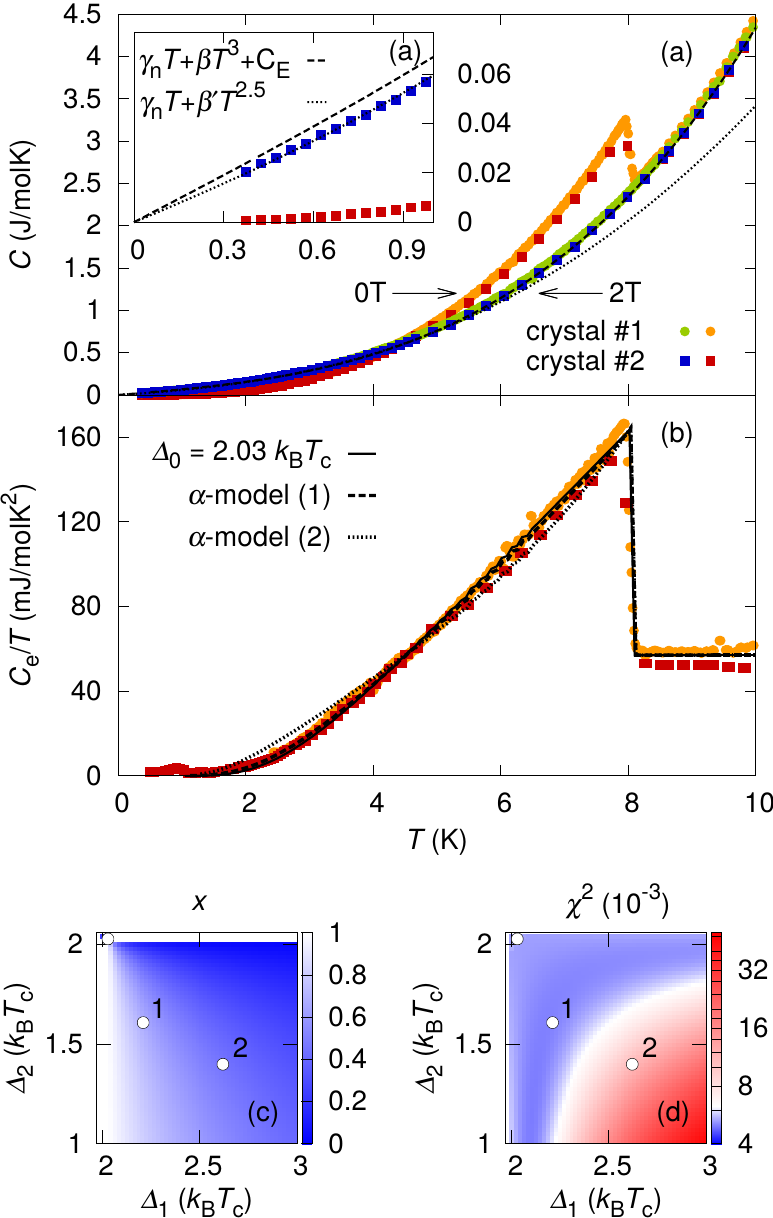}
  \end{center}
  \caption{Temperature dependence of the specific heat. (a) $C(T)$ is shown for two different crystals both at zero field and above the upper critical field $\Hcc$. A fit of the results at 2\,T with Eq.\ \ref{eqn:LaPt4Ge12_debye_spec_heat} falls on top of the data, except below 4\,K, where the phonon contribution follows 
  $\beta' T^{2.5}$ (see inset). (b) The electronic contribution $C_\mathrm{e}(T)$ to the zero-field curve follows a BCS behavior with $\Delta_0 = 2.03\,\kB \Tc$ (solid line). Both samples show the same results within the measurement uncertainty. (c) and (d) Numerical evaluation of the two gap $\alpha$-model. (c) The parameter $x$ of the model was determined from the specific-heat jump at $\Tc$ for the displayed parameter range of $\Delta_1$ and $\Delta_2$. (d) The variance compared to the data in (b), $\chi^2$, reveals a large parameter region around 
  $\Delta_1,\Delta_2 = \Delta_0$ (blue), where the $\alpha$-model is able to describe the data equally well (note the logarithmic scale). The electronic specific heat from the two example parameter sets 1 ($\Delta_1/\kB \Tc=2.21$, $\Delta_2/\kB \Tc=1.61$, $x=0.67$), and 2 ($\Delta_1/\kB \Tc=2.62$, $\Delta_2/\kB \Tc=1.40$, $x=0.45$)  is shown as dashed and dotted curves in (b). Curve 1 falls almost on top of that of the single-band model.}
  \label{fig:C-vs-T}
\end{figure}

The specific heat $C(T)$ of \LaPG\ was measured at zero field and in a field of 2\,T, which is above the superconducting critical field (Fig.\ \ref{fig:C-vs-T}(a)). Our measurements are in good agreement with previous results on polycrystalline samples \cite{gumeniuk_2008}. The data at 2\,T, 
$C_{2\,\mathrm{T}}(T)$, are used as an estimate of the phonon contribution. To obtain the electronic contribution $C_e$ to the specific heat at zero field, the results at 2\,T are subtracted from the zero-field data except for a normal-state electronic contribution $\gamma_\mathrm{n} = C_\mathrm{n}(T)/T = \mathit{const.}$, which is determined by a fit to $C_{2\,\mathrm{T}}(T)$ between $0.3\,\mathrm{K}<T<10\,\mathrm{K}$ using the sum of a Debye and Einstein model,
\begin{equation}
 C(T) = \gamma_\mathrm{n} T + \beta T^3 + C_\mathrm{E}\, ,
 \label{eqn:LaPt4Ge12_debye_spec_heat}
\end{equation}
where $C_\mathrm{E}$ describes the Einstein specific heat from a single phonon mode. The fit result is also plotted in Fig.\ \ref{fig:C-vs-T}(a). From the fit we obtain $\gamma_\mathrm{n} = 56\,\mathrm{mJ\,mol^{-1}K^{-2}}$, 
$\beta = 3.7\,\mathrm{mJ\,mol^{-1}K^{-4}}$ (Debye temperature $\Theta = 208$\,K), and an Einstein temperature of $T_\mathrm{E}=95.5$\,K. The fit describes the data in the whole temperature range reasonably well, only below 4\,K there are small deviations and the phonon contribution follows $\beta'T^{2.5}$. This deviation is most likely due to the complicated phonon spectrum generally observed in filled skutterudites caused by a combination of modes from the filler atom La and modes from the cage structure formed by Pt and Ge \cite{bauer_2008,feldmann_2000,jeon_2016}. The addition of a $T^5$-term is not able to improve the fit. The deviations of the data from the model in Eq.\ \ref{eqn:LaPt4Ge12_debye_spec_heat}, however, can be neglected for the following analysis of the specific heat, but it becomes important when we discuss the thermal conductivity. The result for the electronic contribution to 
$C_{0\,\mathrm{T}}(T)$ is shown in Fig.\ \ref{fig:C-vs-T}(b) as the specific heat coefficient $\gamma(T) = C_\mathrm{e}(T)/T$. We note that the feature in $C_\mathrm{e}/T$ below 1\,K is an artefact due to the subtraction of the lattice contribution to the specific heat.

The specific heat exhibits a sharp superconducting transition at 8.0\,K with a width of 0.13\,K, which is an indication of the good quality of our crystals. The jump height is larger than expected from the predictions of the weak-coupling BCS model ($\Delta_0 = 1.76\,\kB\Tc$), which cannot reproduce our data. However, we are able to describe $C_\mathrm{e}(T)/T$ adjusting the gap to $\Delta_0 = 2.03\,\kB\Tc$. This value is in resonable agreement with the results from photoelectron spectroscopy: $\Delta_0 = 1.95\,\kB T_c$  \cite{nakamura_2012}, and NMR: $\Delta_0 = 1.92\,\kB \Tc$ \cite{kanetake_2010}.

Since superconductivity in other skutterudites is discussed in terms of multiband superconductivity
\cite{zhang_2013,seyfarth_2006,hill_2008,boschenek_2012} and there are results in favor of this interpretation for \LaPG\ as well \cite{zhang_2015_2}, we also consider this possibility based on our specific-heat data. In the case of two gaps, the specific heat might be described by a weighted sum of the contributions from both gaps within the standard two-gap $\alpha$-model \cite{bouquet_2001}
\begin{equation}
 C_{\mathrm{e},\alpha} = xC_\mathrm{e,1} + (1-x)C_\mathrm{e,2}\, .
 \label{eqn:alpha}
\end{equation}
Both $C_\mathrm{e,1}$ and $C_\mathrm{e,2}$ are calculated within the BCS theory, but with variable gap sizes $\Delta_1$ and $\Delta_2$. The third free parameter of this model is the weighting factor $x$. Since our data can already be described with a single gap 
$\Delta_0$, one of the two gaps considered for the $\alpha$-model must be greater, the other one smaller than $\Delta_0$. We therefore evaluate the $\alpha$-model numerically in the parameter range $(1\kB\Tc \lesssim \Delta_1 \lesssim \Delta_0)$, $(\Delta_0 \lesssim \Delta_2 \lesssim 3\kB\Tc)$ using the data between $2\,\mathrm{K}<T<\Tc$.  Using Eq.\ \ref{eqn:alpha}, we determined the corresponding $x$ from the height of the specific heat jump at $\Tc$, which we obtained from the single band calculation with $\Delta_0 = 2.03\,\kB\Tc$. The results for $x$ are shown in Fig.\ \ref{fig:C-vs-T}(c). 

In order to find all parameter sets $(\Delta_1,\Delta_2,x)$, which are able to describe our data, we evaluate the variance
\begin{equation}
 \chi^2 = \frac{1}{n} \frac{1}{T^2\gamma_\mathrm{n}^2} \sum_i^n \left(C_\mathrm{e,\alpha} - C_\mathrm{e} \right)^2\, ,
\end{equation}
which we plot in Fig.\ \ref{fig:C-vs-T}(d) on a logarithmic scale. In this plot, we can identify the single band case in the corner where $\Delta_1,\Delta_2 = \Delta_0$. It also extends vertically and horizontally from this point, where we find either $x=1$ or $x=0$, respectively. However, there is also a large region away from the single-band case, which has the same low variance. As an example, the electronic specific heat calculated from parameter set 1 (see Fig.\ \ref{fig:C-vs-T}(b)) lies almost on top of the single-band calculation. We obtain such a shape and large size of this low-variance region, because a single band is already capable to describe our data. Away from the low variance region, the results from the $\alpha$-model start to deviate from our data, which is demonstrated by the curve from parameter set 2 in Fig.\ \ref{fig:C-vs-T}(b). In summary, our results from the analysis with the $\alpha$-model reveal two possible scenarios: (a) two gaps close to $\Delta_0$ or (b) 
$\Delta_1 \approx \Delta_0$ and $\Delta_2 \neq \Delta_0$ as long as $(1-x)$ stays small. The very good agreement with a single $s$-wave gap excludes the possibility to describe $C_e(T)$ of \LaPG\ using a nodal gap structure almost completely. The question whether there are multiple $s$-wave gaps with different gap sizes involved may ultimately be solved by measurements on the substitution series 
(La$_x$Pr$_{1-x}$)Pt$_4$Ge$_{12}$.


\subsection{$H$-dependence of the specific heat}

\begin{figure}
  \begin{center}
    \includegraphics[width=\linewidth]{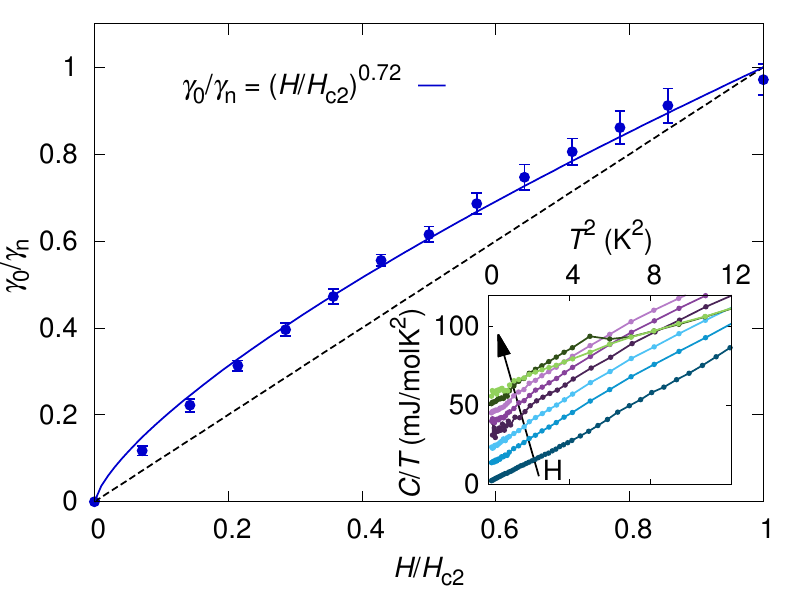}
  \end{center}
  \caption{Magnetic-field dependence of the specific heat. $\gamma_0(H/\Hcc)$ normalized to its normal state value $\gamma_\mathrm{n}$ deviates from a linear behavior (dashed line) and follows a power law (solid line) with an exponent of 0.72. $\gamma_0(H/\Hcc)$ was extracted from a fit to $C(T)/T$ at different fields. A selection of these data is shown in the inset as function of $T^2$ for the equally spaced fields $H=0$, 0.2\,T, ..., 1.4\,T$=\mu_0\Hcc$. The lines are a guides to the eye.}
  \label{fig:C-vs-H}
\end{figure}

The magnetic-field dependence of the $\gamma_0$-coefficient in the superconducting state was determined by an extrapolation of $C_\mathrm{e}(T,H=\mathrm{const})/T$ data to zero temperature assuming an exponential behavior at low temperatures in the superconducting state. Figure \ref{fig:C-vs-H} clearly shows, that $\gamma_0(H/\Hcc)$ normalized to its normal state value $\gamma_{\mathrm{n}}$ follows a sublinear curve $\gamma_0/\gamma_\mathrm{n} = (H/\Hcc)^\eta$ with an exponent of $\eta=0.72$ between $0\leq H\leq \Hcc$. $\mu_0\Hcc=1.4$\,T has been determined from thermal conductivity, see below. This exponent lies in between expectations for nodal superconductors in simplified models ($\eta=0.5$) \cite{volovik_1993} and a fully gapped isotropic superconductor ($\eta=1.0$) \cite{fetter_1969}. However, these predictions take only certain quasiparticle contributions into account: for an $s$-wave superconductor it is the contribution to the density of states from the states within the vortex cores; for nodal superconductors it is a contribution from the delocalized quasiparticles around the nodes. Deviations are, therefore, expected, if one takes further contributions into account. In addition to core states and delocalized quasiparticles, respectively, this would be vortex lattice contributions \cite{ichioka_1999}. 

Since the temperature dependence of the specific heat does not indicate the presence of nodes in the superconducting gap function, we will focus here on the case of fully gaped superconductors. A deviation of $\gamma_0(H)$ from the predicted linear dependence was observed in several $s$-wave superconductors \cite{sonier_1999,nohara_1999,hedo_1998,ramirez_1996,nohara_1997}. Several possible reasons have been suggested: (1) multiband superconductivity; (2) anisotropic Fermi surface; (3) non-linear contributions from the vortex lattice, which arise due to a field-dependent vortex-core radius \cite{sonier_1999,ichioka_1999}. 

We first discuss case (1). In the multiband superconductor MgB$_2$, two distinct field scales were identified \cite{bouquet_2001,boaknin_2003}, which correspond to the two different upper-critical fields of the two gaps. Interestingly, the field dependence of $\gamma_0/\gamma_{\mathrm{n}}$ tracks the field dependence of the thermal conductivity $\kappa_0/\kappa_n$ with the same characteristic field scales. This is not the case in \LaPG, as $\kappa_0/\kappa_n$ shows a curvature with opposite sign (see below) and makes, therefore, a scenario with two gaps of very different size, as in \PrPG, rather unlikely.

\LaPG\ has a complex, anisotropic Fermi surface with several bands crossing the Fermi level \cite{rosner}. This might explain the observed sublinear behavior in $\gamma_0(H)$. A similar scenario has been suggested for the borocarbide superconductor LuNi$_2$B$_2$C \cite{nohara_1997,metlushko_1997}.

We cannot exclude that a field dependent vortex core radius provides a further mechanism for the observed deviations, however, so far there are no indications for such a scenario in \LaPG.


\subsection{$T$-dependence of the thermal conductivity}

\begin{figure}
  \begin{center}
    \includegraphics[width=\linewidth]{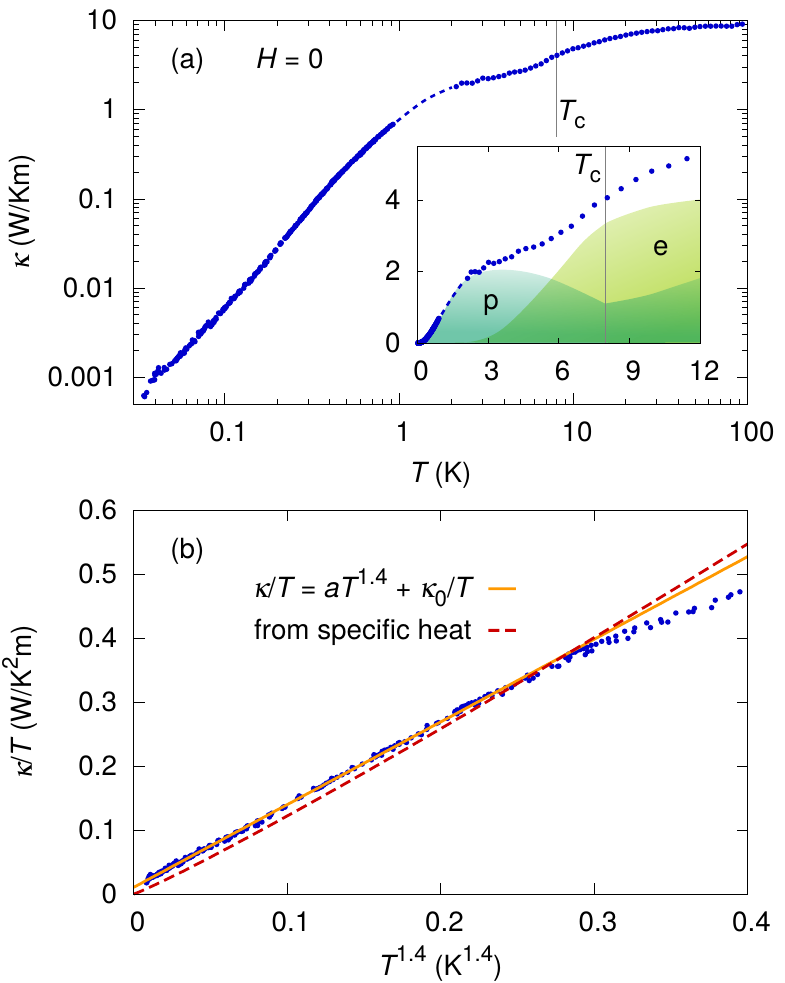}
  \end{center}
  \caption{Temperature dependence of the thermal conductivity at $H=0$. (a) In the double-logarithmic representation up to 100\,K, a clear but smooth drop of $\ka(T)$ at $\Tc$ is visible. A hump appears at lower $T$. Inset: $\ka(T)$ on a linear scale. For an $s$-wave superconductor, one expects a rapidly decreasing $\kaSC{e}{}$ (e) below $\Tc$, while 
  $\kaSC{p}{}$ (p) increases forming a hump. The sketched $T$-dependencies for 
  $\kaSC{e}{}$ and $\kaSC{p}{}$ are inspired by the theory of J.\ Bardeen, G.\ Rickayzen, and L.\ Tewordt (BRT) for $s$-wave superconductors. The dashed lines are guides to the eye. (b) The low-temperature thermal conductivity can be fitted with a power law $\kappa(T)/T \propto T^{1.4}$ (solid yellow) leading to a residual term of 
  $\kappa(T)/T(T\rightarrow 0) = 0.01\,\mathrm{WK^{-2}m^{-1}}$. A second fit (dashed red) with Eq.\ \ref{eqn:LaPt4Ge12_kinetic_phonon_kappa} using the experimental specific-heat results indicates a dominant phonon contribution scattered by sample boundaries.}
  \label{fig:kappa-vs-T}
\end{figure}

Figure \ref{fig:kappa-vs-T}(a) presents the thermal conductivity $\kappa(T)$ at zero field for temperatures up to 100\,K. Figure \ref{fig:kappa-vs-T}(b) shows the section below about 0.5\,K as a function of $T^{1.4}$, which we will discuss first in more detail. The data follow a power law, which we fit with $\kappa(T)/T = \kappa_0/T+bT^a$ below $T=0.4\,\mathrm{K}$. From the fit, we obtain a residual term 
$\kappa_0/T = 0.01\,\mathrm{WK^{-2}m^{-1}}$ ($\kappa_0/\kappa(\Hcc) = 1\%$), which is of the order of our measurement uncertainty. A sizable residual term is expected for a nodal superconductor \cite{graf_1996,durst_2000,shakeripour_2009} due to pair-breaking impurities. Our observed $\kappa_0/T$ is small compared to these expectations and experimental results for unconventional superconductors, \eg\ the $d$-wave superconductors Tl$_2$Ba$_2$CuO$_{6+\delta}$ with $\kappa_0/\kappa(\Hcc) \approx 35\%$ \cite{proust_2002}, and CeIrIn$_5$ with $\kappa_0/\kappa(\Hcc) \approx 20\%$ \cite{shakeripour_2009}. Hence, this result points towards a superconductor with a finite gap everywhere on the Fermi surface. 

The exponent we obtain from the power law fit is $a=1.4$. In an $s$-wave superconductor, the thermal conductivity at sufficiently low $T/\Tc$ is expected to be entirely due to phonons. With the relation
\begin{equation}
 \kappa_\mathrm{p} = \frac{1}{3}C_\mathrm{p}v_\mathrm{p}l_\mathrm{p}\, ,
 \label{eqn:LaPt4Ge12_kinetic_phonon_kappa}
\end{equation}
one can estimate the phonon contribution $\kappa_\mathrm{p}$ from the specific heat $C_\mathrm{p}$ and the mean free path $l_\mathrm{p}$. $C_\mathrm{p}=\beta T^3$ is the phonon specific heat from a Debye model determined above. The sound velocity 
$v_\mathrm{p}$ can be calculated from $C_\mathrm{p}$ using the two equations
\begin{equation}
 \beta T^3= \frac{12\pi^4}{5} N k_B \frac{T^3}{\Theta^3} \quad \mathrm{and} \quad v_\mathrm{p} =
 \frac{k_B\Theta}{\hbar} \left( \frac{V}{6\pi^2 N} \right)^{1/3}\, ,
\end{equation}
where $\Theta$ is the Debye temperature, $V = 7.25\cdot 10^{-9}\mathrm{m^3}$ is the volume of sample \#2 and $N=17 mN_\mathrm{A}/M = 3.84\cdot10^{20} $ is the number of atoms in the crystal. This leads to $\Theta = 208$\,K and $v_\mathrm{p} = 1860$\,m/s. However, the experimental specific-heat data on LaPt$_4$Ge$_{12}$ follow only approximately a $T^3$ dependence, but can be better described by a $T^{2.5}$ law below 4\,K (see inset of Fig.\ \ref{fig:C-vs-T}(a)). Interestingly, the thermal conductivity follows a $T^{2.4}$ dependence at low $T$, which is close to the experimentally determined $T$-dependence of $C_\mathrm{p}$. The almost identical exponents hint at a scenario, where phonons scattered on boundaries are the main contribution to $\kappa$ at low $T$. A fit of the thermal conductivity with an adjusted power law $C_{\mathrm{p}} = \beta' T^{2.5}$, with the velocity $v_\mathrm{p}$ calculated above, and $l_\mathrm{p}$ as free parameter is shown in Fig.\ \ref{fig:kappa-vs-T}(b). It leads to a mean free path 
of $l_\mathrm{p}=0.055$\,mm, which is a 
reasonable value considering the shortest of the sample dimensions 
($0.04 \times0.50 \times1.92$)\,mm$^3$. This nice agreement suggests first boundary scattering as the main contribution to $l_\mathrm{p}$ and a negligible contribution of specular reflections, which would lead to a considerably larger $l_\mathrm{p}$. Secondly, it points towards a negligible electronic contribution to $\kappa$ at low $T$ and thus to a finite gap at every $\boldsymbol{k}$ point of the Fermi surface.

With the results from the low-temperature thermal conductivity, we can also understand the behavior in the whole temperature range (Fig.\ \ref{fig:kappa-vs-T}(a)). At $\Tc$, a clear drop is visible followed by a hump at slightly lower $T$. In general, the thermal conductivity both in the normal state, 
$\kaSC{}{n}$, and in the superconducting state, 
$\kaSC{}{s}$, consists of a phonon and an electron contribution
\begin{eqnarray}
 \kaSC{}{n} &=& \kaSC{e}{n} + \kaSC{p}{n}\, ,\\
 \kaSC{}{s} &=& \kaSC{e}{s} + \kaSC{p}{s}\, .
\end{eqnarray}
Assuming an $s$-wave superconductor, the drop can be attributed to the decreasing number of electronic heat carriers, hence, $\kaSC{e}{s}$ decreases rapidly. The hump indicates an enhanced mean free path of phonons due to the decreasing number of electronic scattering centers, therefore, $\kaSC{p}{s}$ increases below $\Tc$. J.\ Bardeen, G.\ Rickayzen, and L.\ Tewordt (BRT) developed a standard theory for thermal conductivity in the case of an $s$-wave superconductor \cite{bardeen_1959}. In the spirit of the BRT theory, we sketched the temperature dependence for both $\kaSC{p}{}$ and $\kaSC{e}{}$ in the inset of Fig.\ \ref{fig:kappa-vs-T}(a), which can qualitatively explain our observed temperature dependence of the thermal conductivity.


\subsection{$H$-dependence of the thermal conductivity}

\begin{figure}
  \begin{center}
    \includegraphics[width=\linewidth]{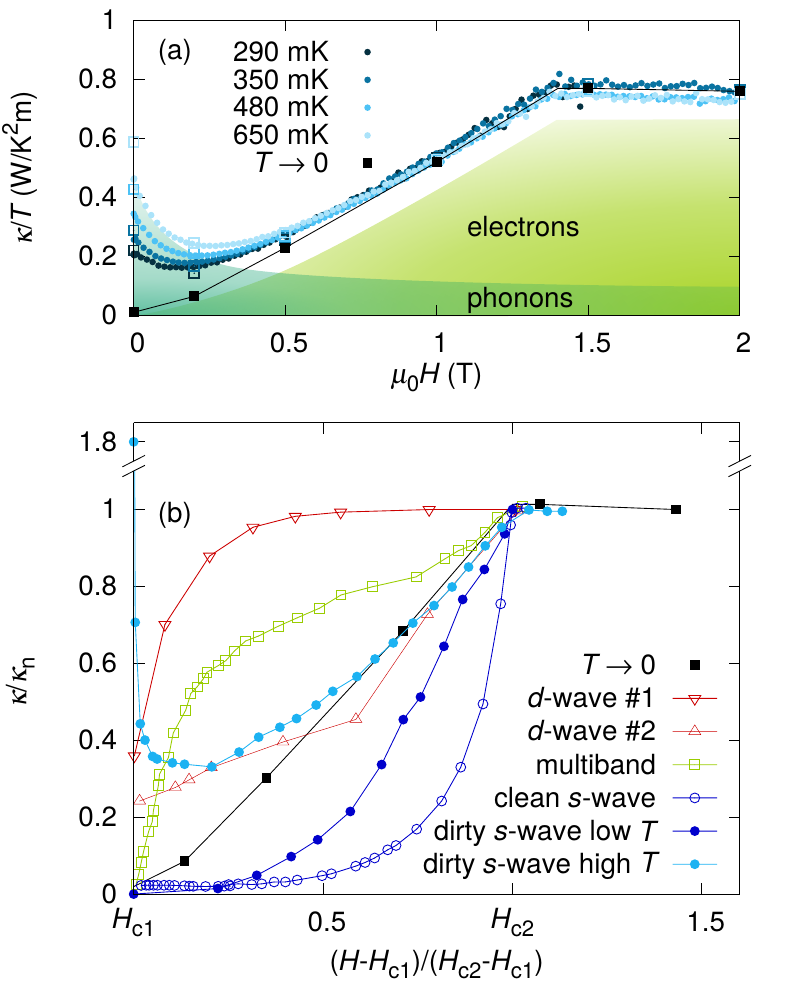}
  \end{center}
  \caption{Magnetic-field dependence of the thermal conductivity. (a) Dots represent field sweeps after zero-field-cooling. Open squares are extracted from temperature sweeps after field-cooling, filled squares are their zero-temperature extrapolations. The line is a guide to the eye. $\kappa(H)/T$ is almost linear and develops a minimum at low fields which becomes more pronounced with higher temperature.  The shaded areas illustrate qualitatively how the electronic and the phonon contribution change with field \cite{dubeck_1963,pesch_1974}. (b)  We compare the field-dependent thermal conductivity data for $T\rightarrow 0$ with the following materials representing different gap symmetries.
  $d$-wave \#1: Tl$_2$Ba$_2$CuO$_{6+\delta}$ ($T/\Tc \rightarrow 0$) \cite{proust_2002}, 
  $d$-wave \#2: CeIrIn$_5$ ($T/\Tc \rightarrow 0$) \cite{shakeripour_2009}, 
   multiband: MgB$_2$ ($T/\Tc = 0.02$) \cite{sologubenko_2002},
  clean $s$-wave: Nb ($T/\Tc = 0.22$) \cite{lowell_1970},
  dirty $s$-wave low $T$: InBi ($T/\Tc = 0.10$) \cite{willis_1976}
  dirty $s$-wave high $T$: Ta$_{80}$Nb$_{20}$ ($T/\Tc = 0.35$) \cite{lowell_1970}.
  Note the different scaling of the field axis to account for the different $H_\mathrm{c1}$ and $H_\mathrm{c2}$.
}
  \label{fig:kappa-vs-H}
\end{figure}

We now turn to the field-dependence of $\kappa(H)/T$ for \LaPG, which is presented in Fig.\ \ref{fig:kappa-vs-H}(a) for selected temperatures between 290\,mK and 650\,mK together with a zero-temperature extrapolation. The measurements were performed on increasing field. A comparison with measurements during decreasing field showed no hysteresis. The field-dependence is consistent with temperature-dependent measurements at finite fields, which were performed on warming and which do not show any difference between field- and zero-field-cooling. 

A clear kink is visible at $\mu_0\Hcc=1.4\,\mathrm{T}$. This is slightly lower than 1.6\,T from the extrapolation of the specific-heat data on polycrystals obtained at 
$T\geq1.8\,\mathrm{K}$ \cite{gumeniuk_2008} and from resistivity data down to 0.3\,K \cite{zhang_2015_2}. The latter observation is in agreement with resistivity measurements on our samples (not shown), which show a slightly higher $\Hcc$ than from thermal transport and specific heat (Fig.\ \ref{fig:C-vs-H}). This deviation is most likely due to surface effects. Below $\Hcc$, the thermal conductivity shows an almost linear field dependence in the zero-temperature limit. For finite temperatures, a minimum appears at low fields. It shifts to higher fields and becomes more pronounced as the temperature is increased.

Generally, $\ka(H)$ is almost constant below $H_\mathrm{c1}$ 
\cite{lowell_1970, gupta_1972, boaknin_2003}, a regime which we will not consider here since $\mu_0 H_\mathrm{c1} \simeq 14\,\mathrm{mT}$ obtained from magnetization measurements is very small in \LaPG.

However, there are drastic changes above $H_\mathrm{c1}$ due to the properties of the vortex state: In addition to the delocalized quasiparticles due to thermal excitations above the gap, there are quasiparticles from the core region, which lead to an increase of $\kaSC{e}{s}$ 
\cite{lowell_1970,caroli_1965,schmidtbauer_1970, golubov_2011}. Around the vortex core, a supercurrent flows, which decays over a distance roughly equal to the penetration depth $\lambda$. The supercurrent with the velocity $\boldsymbol{v}_\mathrm{S}$ leads to a Doppler shift of the energy of delocalized quasiparticles 
$\epsilon(\boldsymbol{k}) \rightarrow \epsilon(\boldsymbol{k}) - \hbar \boldsymbol{k}\cdot\boldsymbol{v}_\mathrm{S} (H)$  \cite{cyrot_1965, volovik_1993}. This changes the excitation spectrum, increases the density of states, and effectively lowers the gap $\Delta$ for $\boldsymbol{k}$ directions with a component parallel to $\boldsymbol{v}_\mathrm{S}$. This effect also increases $\kaSC{e}{s}$ 
\cite{volovik_1993, kuebert_1998, vekhter_1999b} and becomes important especially for superconductors with nodes in the gap. Scattering of electrons and phonons on the vortex lattice decreases 
$\kaSC{e}{s}$ and 
 $\kaSC{p}{s}$ \cite{lowell_1970,caroli_1965,kuebert_1998}.

The sum of all these effects leads to pronounced and typical differences in $\ka(H)$ for materials with different gap structures. They are all anisotropic with respect to the angle between current and field and also with respect to the angle between field and wave vector \cite{lowell_1970,kuebert_1998,brandt_1967,maki_1967}. In the following we restrict the discussion to our case of a field perpendicular to the current. The field dependence of $\kappa(H)$ can be affected by the quality of the sample as exemplified in Fig.\ \ref{fig:kappa-vs-H}b for the case of an $s$-wave superconductor \cite{lowell_1970,willis_1976,caroli_1965,maki_1967}.

In Fig.\ \ref{fig:kappa-vs-H}(b), we compare our data with results on different materials. The materials chosen for this comparison show a field-dependence of the thermal conductivity typical for their superconducting gap symmetry. The key signatures found in \LaPG\ are the finite slope close to $\Hcc$, no residual term $\kappa_0/\kappa_n$ at zero field, and a minimum at intermediate fields, which is absent for $T\rightarrow 0$ but increases in amplitude with increasing $T$. The curve for \LaPG\ compares best with that of InBi, which is a representative example for dirty s-wave superconductors \cite{lowell_1970,gupta_1972,muto_1968,dubeck_1964,willis_1976}.
The most important difference to the clean $s$-wave case is the infinite slope at $\Hcc$ found in the latter one. We also added data for a second dirty superconductor at higher relative temperature to illustrate the appearance of a minimum at intermediate fields very similar to our observation in \LaPG.

The good agreement of $\ka(H)$ in \LaPG\ with that of a dirty $s$-wave superconductor fits to the specific heat results and the temperature dependence of the thermal conductivity. We find in addition, that our sample is actually in the dirty limit, since the Ginzburg-Landau coherence length 
$\xi = 20\,\mathrm{nm}$ is of the same order as the superconducting mean free path 
$l=60\,\mathrm{nm}$. Here, we used the upper critical field to determine $\xi$ with 
$\mu_0\Hcc = {{\it{\Phi}}_0}/{2\pi \xi^2} = 1.4\,\mathrm{T}$. To calculate $l$, we applied
$ 
l = [\frac{\xi^{-2} - 1.6\cdot 10^{12} \rho_0 \gamma_\mathrm{n} \Tc}{1.8\cdot 10^{24}(\rho_0 \gamma_\mathrm{n} \Tc)^2}]^{0.5}\,\mathrm{cm}\, ,
$ 
\cite{orlando_1979} where we use the electronic specific heat coefficient $\gamma_\mathrm{n} = 1600$ in the unit $\mathrm{erg\,K^{-2}/cm^{-3}}$, the residual resistivity $\rho_0 = 4\cdot 10^{-6}$ in the unit $\mathrm{\Omega cm}$ (from crystal \#3, not shown), $\Tc = 8$ in the unit K, and the aforementioned coherence length $\xi = 2\cdot 10^{-6}$ in the unit cm.

The good agreement with results for a dirty $s$-wave superconductor implies, that the field dependence of the thermal conductivity is composed of a phonon part $\kaSC{p}{s}$, which decreases with increasing field due to scattering on vortices, and an electronic contribution, which increases due to an enhanced number of localized states in the vortex cores tunneling to ever closer neighboring vortices. This behavior is indicated in Fig.\ \ref{fig:kappa-vs-H}(a).

Since our specific-heat results indicate that a description within a two-band model is also consistent with the data, we like to analyze our thermal-transport data with respect to this scenario as well. A typical feature of $\kappa(H)$ in multiband systems is a plateau at intermediate fields as shown in Fig.\ \ref{fig:kappa-vs-H}(b) for MgB$_2$, which is attributed to the suppression of the smaller of the two gaps. We do not observe such change of curvature in \LaPG. However, this change might be very weak as \eg\ in the case of NbSe$_2$ \cite{boaknin_2003}. 

Let us now compare our thermal conductivity results with observations and predictions for the case of nodes in the superconducting gap. There are both theoretical \cite{kuebert_1998,vekhter_1999b} and experimental \cite{watanabe_2004,machida_2012,izawa_2001_sr2ruo4} reports on a minimum in $\kappa(H)$ for unconventional superconductors. The slope above the minimum can vary considerably depending on the type of gap and the purity of the sample \cite{kuebert_1998,vekhter_1999b,won_2001} (\cf\ Fig.\ \ref{fig:kappa-vs-H}(b)). Such a minimum is consistent with our observation. Hence, this property does not rule out nodes in the gap of \LaPG\ completely. However, one would expect a sizable residual term for $T\rightarrow 0$ in a nodal superconductor. In contrast, both single-band and multiband $s$-wave superconductors show an insignificantly small residual term exactly as our results on \LaPG.

\section{Conclusion}

We performed specific-heat and thermal-conductivity measurements to investigate the superconducting order parameter of \LaPG. The specific heat shows a sharp superconducting transition at 8.0\,K and its zero-field temperature-dependence in the superconducting state can nicely be described assuming a single BCS $s$-wave gap of the size $\Delta = 2.0\,\kB\Tc$. The field dependence of the $\gamma_0$-coefficient deviates from the simple linear behavior for $s$-wave pairing. Such a deviation can arise, if the Fermi surface shows strong anisotropies as is the case for \LaPG. We used the specific-heat results to analyze the low-temperature thermal-transport in detail. The temperature dependence of the thermal conductivity reveals a dominant phonon contribution at low temperatures and a negligible electronic residual term. Together with the dependence of the thermal conductivity on magnetic field, this behavior suggests a single gap BCS $s$-wave superconductor.

However, we cannot completely exclude a two-band scenario from our experimental results, in particular if both gaps have a similar size $\Delta/k_\mathrm{B}T_c$ or if one of them has only a tiny contribution to the thermal-transport and thermodynamic properties. The analysis of the specific heat data enables us to limit the possible range of amplitudes $\Delta_1$ and $\Delta_2$ and weighting factors $x$ within the two-gap $\alpha$-model. The possible parameter range is comparatively large, since a single-band model is already sufficient to describe our experimental results.

Since the sister compound \PrPG\ is discussed to be a multigap and/or unconventional superconductor, further thermal conductivity and specific heat measurements on the substitution series (Pr$_{1-x}$La$_x$)Pt$_4$Ge$_{12}$ are highly desired. They might help to resolve the question, if \PrPG\ and \LaPG\ have compatible superconducting order parameters as was suggested by the smooth evolution of $T_c(x)$ across the substitution series.

\section*{Acknowledgement}

We are indebted to R.\ Daou and H.\ Rosner for valuable discussions.


%

\end{document}